\def\<{\langle}
\def\>{\rangle}
\documentclass[prl,aps,tightenlines,twocolumn,showpacs]{revtex4}
\usepackage{epsfig}

\begin{document}
\title{Maximizing the entanglement of two mixed qubits}

\author{W.~J. Munro$^{1,2}$, D. F. V. James$^{3}$, A. G. White$^{1}$, and P. G.
Kwiat$^{4}$}

\affiliation{$^{1}$ Special Research Centre for Quantum Computer
Technology, University of Queensland, Brisbane, AUSTRALIA}

\affiliation{$^{2}$  Hewlett-Packard Laboratories, Filton Road,
Stoke Gifford, Bristol BS34 8QZ, UK}

\affiliation{$^{3}$ Theory Division, T-4, Los Alamos National
Laboratory, Los Alamos, New Mexico, USA}


\affiliation{$^{4}$ Dept. of Physics, University of Illinois,
Urbana-Champaign, Illinois, USA.}

\affiliation{LA-UR 01-0991}

\date{\today}

\begin{abstract}
Two-qubit states occupy a large and relatively unexplored Hilbert
space. Such states can be succinctly characterized by their degree
of entanglement and purity. In this letter we investigate
entangled mixed states and present a class of states that have the
maximum amount of entanglement for a given linear entropy.
\end{abstract}


\maketitle

With the recent rapid developments in quantum information there
has been a renewed interest in multiparticle quantum mechanics and
entanglement. The properties of states between the pure,
maximally-entangled, and completely mixed (separable) limits are
not completely known and have not been fully characterized.  The
physically allowed degree of entanglement and mixture is a timely
issue, given that entangled qubits are a critical resource in many
quantum information applications (such as quantum
computation\cite{DiVincenzo95,Vedral98}, quantum
communication\cite{Schumacher96}, quantum cryptography\cite{Ekert
91,Naik00} and teleportation\cite{Bennett93,Bouwmeester97}), and
that entangled mixed states could be advantageous for certain
quantum information situations\cite{Cleve99}.

The simplest non-trivial multiparticle system that can be
investigated both theoretically and experimentally consists of
two-qubits. A two-qubit system displays many of the paradoxical
features of quantum mechanics such as superposition and
entanglement.  Extreme cases are well known and clear enough:
maximally-entangled two particle states have been produced in a
range of physical systems \cite{Aspect
82,Kwiat99,Turchette98,Turchette95}, while two-qubits have been
encoded in product (non-entangled) states \cite{Braunstein99} via
liquid NMR \cite{Chuang98}.  Recently, however White {\it et al.}
have experimentally generated polarization-entangled photons in
both non-maximally entangled states\cite{White99}, and general
states with variable degree of mixture and entanglement
\cite{White00}.

In this letter we explore theoretically the domain between pure,
highly entangled states, and highly mixed, weakly entangled
states.  We will partially characterise\cite{partial} such
two-qubit states by their {\it purity\/} and {\it degree of
entanglement\/} \cite{Bennett96}. Specifically, we address the
question: What is the form of maximally-entangled mixed states,
that is, states with the maximum amount of entanglement for a
given degree of purity? Ishizaka {\it et al.} \cite{Ishizaka00}
have proposed two-qubit mixed states in which the degree of
entanglement cannot be increased further by any unitary operations
(the Werner state\cite{Werner89} is one such example). A numerical
exploration of the entanglement - purity plane is used to
establish an upper bound for the maximum amount of entanglement
possible for a given purity, and vice versa.  We derive an
analytical form for this class of {\it maximally-entangled mixed
states} (MEMS)  and show it to be optimal for the entanglement and
purity measures considered.

Currently a variety of measures are known for quantifying the
degree of entanglement in a bipartite system.  These include the
entanglement of distillation\cite{Bennett96}, the relative entropy
of entanglement\cite{Vedral98}, but the canonical measure of
entanglement is called the {\it entanglement of formation} (EOF)
\cite{Bennett96} and  for an arbitrary two-qubit system is given
by\cite{Coffman99}
\begin{eqnarray}\label{EOF}
E_F(\hat{\rho})=h\left(\frac{1+\sqrt{1-\tau}}{2}\right),
\end{eqnarray}
where $h(x) = -x\log_2(x) - (1 - x)\log_2(1 - x)$ is Shannon's
entropy function and $\tau$, the ``tangle'' \cite{Coffman99}
(``concurrence'' squared) is given by
\begin{eqnarray}
\tau={\cal
C}^{2}=\left[\max\{\lambda_1-\lambda_2-\lambda_3-\lambda_4,0\}\right]^{2},
\end{eqnarray}
Here the $\lambda$'s are the square roots of the eigenvalues, in
decreasing order, of the matrix,
%
$\hat{\rho} \tilde{\hat{\rho}} = \hat{\rho}\;\sigma_{y}^{A}
\otimes \sigma_{y}^{B} \hat{\rho}^{*} \sigma_y^A \otimes
\sigma_{y}^{B}$,
%
where $\hat{\rho}^{*}$ denotes the complex conjugation of $\hat{\rho}$
in the computational basis $\{|00\>, |01\>,|10\>,|11\>\}$, and is an
anti-unitary operation. Since the
entanglement of formation $E_F$ is a strictly monotonic function of
$\tau$, the maximum of $\tau$ corresponds to the maximum of $E_F$.
Thus in this letter we use the tangle directly as our measure of
entanglement. For a maximally-entangled pure state $\tau=1$, while for
a unentangled state $\tau=0$.

There exist for the degree of mixture of a state a number of
measures.  These include the von Neumann entropy of a state, given
by $\rm{S}=-\rm{Tr}[\hat{\rho} \rm{ln}\hat{\rho}]$
\cite{vonNeumann55}, and the purity ${\rm Tr} [\hat{\rho}^{2}]$.
In this letter we use the linear entropy given by\cite{Bose00}
\begin{eqnarray}
{\cal S}_{L}= {4 \over 3} \left\{1-{\rm Tr}\left[
\hat{\rho}^{2}\right]\right\}.
\end{eqnarray}
which ranges from 0 (for a pure state) to 1 (for a maximally-mixed
state). The linear entropy is generally a simpler quantity to
calculate and hence its choice here.

\begin{figure}[!h]
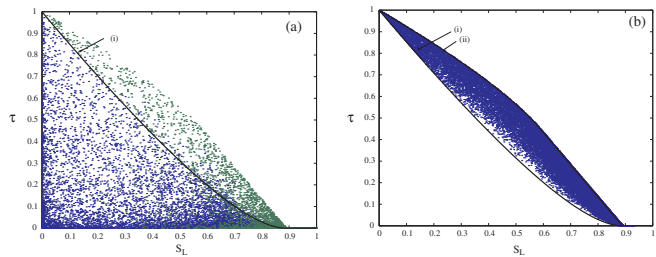

\center{\epsfig{figure=Fig1a.epsf,width=45mm}\epsfig{figure=Fig1b.epsf,width=41mm}}
\caption{Plot of the tangle, $\tau$, and linear entropy, ${\cal
S}_{L}$, of numerically generated two-qubit random matrices.  Two
sets of data are plotted: a) 30,000 randomly generated matrices,
which show the extent of physical states in the
entanglement-purity plane; b) 30,000 randomly generated matrices
weighted to explore the boundary region. Also shown are analytical
curves for: i) the Werner state, a mixture of the
maximally-entangled state and the maximally-mixed state; and ii)
the maximally-entangled mixed states, states with the maximal
amount of entanglement for a given degree of linear entropy (or
vice versa).  See text for further details.} \label{fig1}
\end{figure}

Let us now examine our two-qubit states and the region they occupy
in the tangle-linear entropy plane.  We begin by randomly
generating two million density matrices representing physical
states, and determining their linear entropy and tangle.  In Fig.\
(\ref{fig1}a) we display a subset of these results for thirty
thousand points.  We see that quite a large region of this plane
is filled with physically acceptable states (obviously a
maximally-mixed, maximally-entangled, state is not possible).
Zyczkowski {\it et. al.}\cite{Zyczkowski 99} have  preformed
similar numerical studies but their work focused on how many
entangled states are in the set of all quantum states.

In Fig.\ (\ref{fig1}a) we have also explicitly plotted the tangle
versus linear entropy for the Werner state, a mixture of the
maximally-entangled state and the maximally-mixed state
\cite{Werner89}:
\begin{eqnarray}\label{werner}
\hat{\rho} = {{1-\gamma}\over{4}} I_{2}\otimes I_{2} + \gamma
|\Phi_{+}\>\< \Phi_{+}|,
\end{eqnarray}
where $I_{2}$ is the identity matrix and $|\Phi_{+}\>={1\over
\sqrt 2} \left[|0\>|0\>+|1\>|1\> \right]$. We have labelled our
orthogonal qubit states by $|0\>$ and $|1\>$.  This Werner state
is entangled (inseparable) for $\gamma>1/3$\cite{Bennett96a} and
maximally-entangled when $\gamma=1$.  The results from Fig.\
(\ref{fig1}a) clearly indicate a class of states that have a
larger degree of entanglement for a given linear entropy than the
Werner states.  We also generated a second set of data (by random
perturbations about the maximally-entangled mixed states) so as to
examine the boundary of possible states, which in the previous
data set was a sparsely populated region.  As can be seen in Fig.\
(\ref{fig1}b), a definite boundary to the physically possible
states exists.

Let us now analytically determine the form of these
maximally-entangled mixed states (MEMS).  As our starting point, let
us consider the Werner state given by (\ref{werner}).
How can one increase its degree of entanglement
without changing its purity, or, alternatively, how can one increase
its linear entropy given a certain degree of entanglement?  In
deriving our ansatz we will note the following points:

\begin{itemize}

\item In the Werner state (\ref{werner}) all the entanglement
arises from the $\gamma |\Phi_{+}\>\< \Phi_{+}|$ term, and hence,
to leave the degree of entanglement fixed while increasing the
linear entropy this term needs to remain untouched. Local unitary
operations will not affect the degree of entanglement or linear
entropy.

\item The $I_{2}\otimes I_{2}$ term of the Werner states comprises the
maximally-mixed state.  It can be written as an equal incoherent
mixture of the four Bell states
\begin{eqnarray}
|\Psi_{\pm}\>&=&{1\over \sqrt 2} \left[|0\>|1\>\pm|1\>|0\> \right],\\
|\Phi_{\pm}\>&=&{1\over \sqrt 2} \left[|0\>|0\>\pm|1\>|1\> \right],
\end{eqnarray}
In our ansatz, if we increase the amount of any of the
$|\Psi_{\pm}\>$ or $|\Phi_{-}\>$ Bell states, then the net
entanglement in the total system generally decreases.

\item In a general two-qubit density matrix there are two types of
off-diagonal terms, those that represent the entanglement and those
that represent single particle superposition.  These
single particle superposition terms can be set to zero by
local linear operations, and so, by definition, cannot change
the net entanglement.

\item The diagonal elements of the two-qubit density matrix do not
affect the system's maximum entanglement (given a specified amount of
$|\Phi_{+}\>\< \Phi_{+}|$.  The diagonal elements,
however, have a significant impact on the linear entropy.

\end{itemize}
These principles lead us to postulate an ansatz of the form
\begin{eqnarray}\label{ansatz}
\hat{\rho}=\left(
\begin{array}{cccc}
x+{\gamma\over 2}&  0  & 0  & {\gamma\over 2} \\
0 & a & 0 &  0 \\
0 & 0 & b & 0 \\
{\gamma\over 2} &  0  & 0  & y +{\gamma\over 2}\\
\end{array}
\right).
\end{eqnarray}
This comprises a mixture of the maximally-entangled Bell state
$|\Phi_{+}\>$ and a mixed diagonal state (who populations are
specified by the real  and non-negative parameters a,b,x,y).
%
%
%
%
Without loss of generality we choose $\gamma$ to be a positive
real number, which ensures that the ansatz density matrix is
positive semi-definite. From normalization,
\begin{eqnarray}\label{traceconst}
x+y+a+b+\gamma=1,
\end{eqnarray}
The linear entropy is simply given by
\begin{eqnarray}
{\cal S}_{L}= {4 \over 3}
\left\{1-a^{2}-b^{2}-x^{2}-y^{2}-\gamma
\left(x+y\right)-\gamma^{2} \right\}
\end{eqnarray}
with the concurrence given by
\begin{eqnarray}
{\cal C}&=&{\rm Max}\left[ \gamma- 2 \sqrt{a b},0 \right].
\end{eqnarray}

To determine the form of the two-qubit maximally-entangled mixed
states, we begin by  specifying that the concurrence ${\cal C}$
must be greater than zero. Thus ${\cal C}={\rm Max}\left[\gamma- 2
\sqrt{a b},0 \right] = \gamma- 2 \sqrt{a b}\geq 0$ and therefore
is maximized when ${\cal C}=\gamma$. This requires either $a=0$
and/or $b=0$ (without loss of generality we set $b=0$).
Using the normalization constraint given by (\ref{traceconst}),
the linear entropy is given by
\begin{eqnarray}\label{SL1}
{\cal S}_{L}= {4 \over 3}
\left\{   2 a+ \left(\gamma+2 x\right)\left(1-a-\gamma \right)
-2 x^{2}-2 a^{2}\right\}.
\end{eqnarray}
Calculating the turning point of (\ref{SL1}) we find that $\partial
{\cal S}_{L} / \partial x=0$ when either $x=0$ (a minimum) or $2
x=1-a-\gamma$ (a maximum) and $\partial {\cal
S}_{L}/\partial a=0$ when either $a=0$ (a minimum) or $4 a=2-2
x-\gamma$ (a maximum).  First examining the
$\partial {\cal S}_{L}/\partial x$ stationary solution and the
maximum given by $2 x=1-a-\gamma$, we observe that this condition requires
$x=y$.  If $a=1-\gamma$ then the stationary point
corresponds to a turning point.  We now need to examine several
parameter regimes to determine the optimal solution.  The first region
has concurrence values in the region $2/3\leq {\cal C}\equiv \gamma\leq 1$.  In
this region the optimal situation occurs when $x=0$ and $a=1-\gamma$.
This means the maximally-entangled mixed state has the form
\begin{eqnarray}
\hat{\rho}_{\rm\scriptsize MEMS}=\left(
\begin{array}{cccc}
{\gamma/ 2}&  0  & 0  & {\gamma/ 2} \\
0 & 1-\gamma & 0 &  0 \\
0 & 0 & 0 & 0 \\
{\gamma/ 2} &  0  & 0  & {\gamma/ 2} \\
\end{array}
\right).
\end{eqnarray}
The second regime occurs for $0\leq {\cal C}\equiv \gamma \leq 2/3$.
In this case the optimal solution occurs
when $a=1/3$ and $x+\gamma/2=1/3$.
The optimal maximally-entangled mixed state
in this region has the form
\begin{eqnarray}
\hat{\rho}_{\rm\scriptsize MEMS}=\left(
\begin{array}{cccc}
{1/ 3}&  0  & 0  & {\gamma/ 2} \\
0 & {1/ 3} & 0 &  0 \\
0 & 0 & 0 & 0 \\
{\gamma/ 2} &  0  & 0  & {1/3} \\
\end{array}
\right).
\end{eqnarray}
In this case the diagonal elements do not vary with $\gamma$.
Combining both these solutions, we can obtain (up to local unitary
transformations) the following single
explicit form for the maximal entangled mixed state:
\begin{eqnarray}\label{hitchcock}
\hat{\rho}_{\rm\scriptsize MEMS}=\left(
\begin{array}{cccc}
g(\gamma)&  0  & 0  & {\gamma/ 2} \\
0 & 1-2g(\gamma) & 0 &  0 \\
0 & 0 & 0 & 0 \\
{\gamma/ 2} &  0  & 0  & g(\gamma) \\
\end{array}
\right),
\end{eqnarray}
where
\begin{eqnarray}
g(\gamma)=\left\{
\begin{array}{cccc}
\gamma/2 & & &{\cal C}\equiv \gamma \geq 2/3\\
1/3 & & &{\cal C}\equiv \gamma < 2/3
\end{array}
\right.
.
\end{eqnarray}
The degree of entanglement for this maximally-entangled mixed state
is simply $\tau=\gamma^{2}$, while the linear entropy has the form
\begin{eqnarray}
{\cal S}_{L}&=&\frac{2}{3}\left[4\,g(\gamma)\,\left( 2 - 3\,g(\gamma)\right)
-\gamma^2\right].
\end{eqnarray}

In figure (\ref{fig1}) we have plotted the tangle versus the linear
entropy for the Werner state, and the numerically determined
maximally-entangled mixed state.  Our analytic expression for the
state  (\ref{hitchcock}) perfectly overlays the numerically
generated optimal curve.  It is clear that these states have a
significantly greater degree of entanglement for a given linear
entropy than the corresponding Werner states.  The maximally-entangled
mixed state  and Werner state curves join each other at
two points in the tangle -
linear entropy plane.  The first and most obvious point occurs at
$(\tau,S_{L})=(1,0)$ (here both states are maximally entangled).
The second point occurs at $(\tau,S_{L})=(0,8/9)$.
Here the two states are given by,
\begin{eqnarray}
\hat{\rho}_{\rm\scriptsize WERNER}=\left(
\begin{array}{cccc}
{1\over 3}&  0  & 0  & {1\over 6} \\
0 & {1\over 6} & 0 &  0 \\
0 & 0 & {1\over 6} & 0 \\
{1\over 6} &  0  & 0  & {1\over 3}
\end{array}
\right),\;\;
\hat{\rho}_{\rm\scriptsize MEMS}=\left(
\begin{array}{cccc}
{1\over 3}&  0  & 0  & 0 \\
0 & {1\over 3} & 0 &  0 \\
0 & 0 & 0 & 0 \\
0 &  0  & 0  & {1\over 3}
\end{array}
\right).
\end{eqnarray}
Neither state is entangled.  We observe that
$\hat{\rho}_{\rm\scriptsize MEMS}$  at this point has
no nonzero off-diagonal elements but the Werner state does. The
maximally-entangled mixed state is entangled as soon
 as the off-diagonal elements are
nonzero ($\gamma >0$, while the Werner state requires $\gamma>1/3$ to
be entangled). Though $\hat{\rho}_{\rm\scriptsize WERNER}$ and
$\hat{\rho}_{\rm\scriptsize MEMS}$
have different forms they have the same degree of entanglement (zero) and
linear entropy.  Because of the way the
maximally-entangled mixed state has been
constructed, it never attains a linear entropy $S_{L}=1$.  The Werner
state attains this point because of its maximally-mixed component.

To confirm that our analytic solution is optimal
and that no density matrix has a greater degree of entanglement for a given
linear entropy than the state (\ref{hitchcock}), we generated one million
further random density matrices. We found that the
maximally-entangled mixed state is indeed optimal.
It is interesting to note, however, that the  state is
only optimal for mixture measures based on
${\rm Tr}\left[\hat{\rho}^{2}\right]$; if instead the degree of mixture
is measured for instance by the entropy\cite{vonNeumann55}
the state is not optimal.

\begin{figure}[!h]
\center{\epsfig{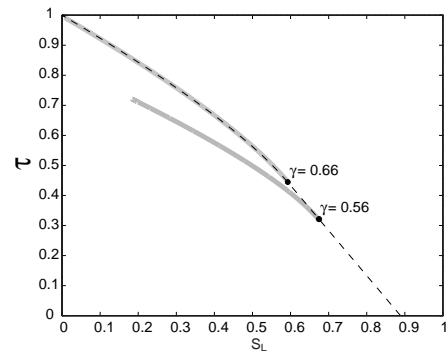}} \caption{ Plot of
the tangle $\tau$ versus linear entropy $S_{L}$ for the
maximally-entangled mixed state  (dotted line). By employing a
concentration protocol[26], an initial state (solid circle), can
be manipulated to produce a range of new states (solid grey lines)
with improved entanglement and linear entropy characteristics. }
\label{fig2}
\end{figure}

Lastly, how does our class of maximally-entangled mixed states
compare with those predicted by Ishizaka and
Hiroshima\cite{Ishizaka00}?  Ishizaka's two-qubit mixed states,
the Werner state being a specific example, were chosen so that the
degree of entanglement of such states cannot be increased further
by unitary operations. In contrast we have derived a class of
states that have the maximum amount of entanglement for a given
linear entropy (and vice versa). Therefore our states are members
of the Ishizaka {\it et al.} class by definition, although they
were not explicitly considered\cite{Ishizaka00}. The Ishizaka {\it
et al.} result indicates that a maximally-entangled mixed state
cannot have the degree of entanglement increased by unitary
operations. This state can however have its entanglement increased
by a simple and experimentally realizable non-unitary
concentration protocol recently proposed by Thew and
Munro{\cite{Thew00}}. Such a protocol is a based on generalization
of the Procrustean method originally introduced for pure
states\cite{Bennett96b} and recently demonstrated
experimentally\cite{Kwiat 00}. In Figure (\ref{fig2})  we display
the results of the concentration protocol for two initial
conditions. The solid curves represent a range of states that are
obtainable, from the maximally-entangled mixed state, as the
concentration protocol is applied to improve the output state
characteristics. We observe that for all $\gamma$ the output
characteristics can be significantly improved (solid grey lines).
In fact for  $\gamma \geq 2/3$  the  maximally-entangled mixed
state can be concentrated up the dashed curve to a
maximally-entangled pure state.

To summarize, we have discovered a class of partially entangled
mixed two-qubit states that have the maximum amount of
entanglement for a given linear entropy.  An analytical form for
these states was derived and they were shown to have significantly
more entanglement for a given degree of purity than the Werner
states.  The properties of these states are still largely unknown
and require significant exploration.  Open questions such as ``can
such states be realized experimentally'', ``to what extent do they
violate Bell inequalities?'', and ``do they have information
processing advantages over other states''   are the subject of
current investigation.\\

We wish to thank K. Nemoto and G. J. Milburn for encouraging
discussions. WJM and AGW would like to acknowledge the support of
the Australian Research Council while DFVJ would like to thank the
University of Queensland for their hospitality during his visit.


\begin{thebibliography}{99}
\bibitem{partial}
Arbitrary two-qubit states are uniquely specified by 15 real
parameters [8]. The degree of entanglement and mixture do not
determine all these 15 parameters absolutely.
\end{thebibliography}
\end{document}